\documentclass[a4paper]{jpconf}
\usepackage{graphicx}
\usepackage{amsmath,amsfonts,amssymb,graphicx,graphics,color}
\usepackage{hyperref}
\usepackage[latin9]{inputenc}
\usepackage{pifont}
\usepackage{subfigure,epsfig,epstopdf}
\usepackage{bm}
\usepackage{enumerate}
\begin{document}

\newcommand{\xip}{\xi_\perp}
\newcommand{\ep}{e_\perp}
\newcommand{\be}{\begin{equation}}
\newcommand{\ee}{\end{equation}}
\newcommand{\bea}{\begin{eqnarray}}
\newcommand{\eea}{\end{eqnarray}}
\newcommand{\mnras}{{Mon.Not.Roy.Astron.Soc.}}
\newcommand{\red}{\textcolor{red}}
\newcommand*{\bigchi}{\mbox{\large$\chi$}}
\def\be{\begin{equation}}
 \def\ee{\end{equation}}
 \def\l{\lambda}
 \def\a{\alpha}
 \def\b{\beta}
 \def\g{\gamma}
 \def\d{\delta}
 \def\e{\epsilon}
 \def\m{\mu}
 \def\n{\nu}
 \def\t{\tau}
 \def\p{\partial}
 \def\s{\sigma}
 \def\r{\rho}
 \def\sl{\ds}
 \def\ds#1{#1\kern-1ex\hbox{/}}
 \def\sla{\raise.15ex\hbox{$/$}\kern-.57em}
 \def\nn{\nonumber}
 \newcommand{\bth}{{\bf 3}}
 \newcommand{\btw}{{\bf 2}}
 \newcommand{\bon}{{\bf 1}}
 \def\Tr{\textnormal{Tr}}
 \def\th{\theta}
 \def\({\left(}
 \def\){\right)}
 \def\[{\left[}
 \def\]{\right]}

\title{Meson physics in asymmetric matter}

\author{Andrea Mammarella}

\address{Laboratori Nazionali del Gran Sasso, via G. Acitelli 22, Assergi (AQ), Italy}

\ead{andrea.mammarella@lngs.infn.it}

\begin{abstract}
This paper describes dynamic and thermodynamic (at $T=0$) properties of mesons in asymmetric matter in the framework of Chiral Perturbation Theory. We consider a mesonic system at vanishing temperature with nonzero isospin chemical potential and strangeness chemical potential and we study their effect on the phase diagram. We also study meson masses and mixing in the resulting normal phase, pion condensation phase and kaon condensation phase. We find differences with previous papers regarding meson masses and mixing in the condensed phases; the results presented here are supported by theory group analysis and direct calculations. Pressure, density and equation of state of the system at $T=0$ and nonzero $\mu_I$ are calculated, finding remarkable agreement with analogue studies performed by lattice calculations.
\end{abstract}

\section{Introduction}
\label{intro}
Understanding the QCD phase diagram is important for many different phenomena, for example the astrophysics of compact stars or heavy-ion collisions. It is known that depending on the value of the isospin chemical potential, $\mu_I$, and on the value of the strangeness chemical potential,  $\mu_S$, three different phases can be realized: the normal phase, the pion condensed ($\pi c$) phase and the kaon condensed ($Kc$) phase~\cite{Migdal:1990vm,Son:2000xc,Kogut:2001id}. The realization of a mesonic condensate can drastically change the low energy properties of matter, including the mass spectrum and the lifetime of mesons.\\

Previous analysis of the meson condensed phases by QCD-like theories were developed in \cite{Kogut:1999iv, Kogut:2000ek}. Pion condensation in two-flavor quark matter was studied in~\cite{Son:2000xc, Son:2000by} and in three-flavor quark matter in~\cite{Kogut:2001id}. In particular,  the phase diagram as a function of $\mu_I$ and $\mu_S$ was presented in~\cite{Kogut:2001id}. Finite temperature effects in  $SU(2)_L \times SU(2)_R$ chiral perturbation theory ($\chi$PT) have been studied in~\cite{Loewe:2002tw, Loewe:2004mu, He:2005nk}. One remarkable property of quark matter with nonvanishing isospin chemical potential is that it is characterized by a real measure, thus the lattice realization can be performed with standard numerical algorithms~\cite{Alford:1998sd, Kogut:2002zg}. The $\pi c$ and the $K c$ phases have  been studied by NJL models in~\cite{Toublan:2003tt, Barducci:2004tt, Barducci:2004nc} and by random matrix models in \cite{Klein:2004hv}. All these models find results in qualitative and quantitative agreement; in particular, the phase diagram of matter has been firmly established. We have studied the masses and mixing of mesonic states and we have found results that disagree with ~\cite{Kogut:2001id} that, up to our knowledge, is the only paper in which they are studied. \\
The thermodynamic properties of the condensed phases have been studied by LQCD simulations in~\cite{Detmold:2012wc, Detmold:2008yn}. Previously, various  results on  the $\pi c$ phase were derived   in~\cite{He:2005nk} by an NJL model. In particular,  the equation of state (EoS) of the NJL model was presented. Recently in~\cite{Andersen:2015eoa,Cohen:2015soa, Graf:2015pyl} a perturbative analysis of QCD at large isospin density has been presented. Those pQCD  results are consistent with LQCD  for $\mu_I \gtrsim 3 m_\pi$~\cite{Graf:2015pyl}, where $m_\pi$ is the pion mass. At smaller values of $\mu_I$ it seems that pQCD  underestimates the energy density and it is not able to capture the condensation mechanism. However, for small values of  $\mu_I$ (and $\mu_S$), $\chi$PT can be used. 

In this article we briefly review how to include chemical potentials in $\chi PT$ \cite{Gasser:1983yg, Leutwyler:1993iq, Ecker:1994gg,  Scherer:2002tk, Scherer:2005ri}, then we describe some phenomenological properties related to this inclusion, like the existence of the different phases already listed and the changes in meson masses and mixing. We will also show how the inclusion of chemical potentials affects meson properties \cite{Mammarella:2015pxa}. Finally we will study the effect of nonzero isospin and strangeness chemical potentials on the pressure, density and equation of state of the system \cite{Carignano:2016rvs}.

The paper is organized as follows. In Sec.~\ref{mod} we describe the model and how it predicts different phases. In Sec.~\ref{mm} we show how group theory tools can be used to calculate the mass eigenstates and the meson mixing in the condensed phases and we list and discuss the results obtained. In sec \ref{term} we study the termodynamic properties of the system. Finally, in Sec. ~\ref{conc}, we summarize the results.

\section{Model} \label{mod}
\subsection{Lagrangian and definitions}
In this section we briefly review the  model that we are going to use in the following. It is the one described in \cite{Kogut:2001id}. It is a chiral effective Lagrangian at the lowest order in the momenta.
The general ${\cal O}(p^2)$ Lorentz invariant Lagrangian density describing the pseudoscalar mesons can be written as \cite{Gasser:1983yg}
\be\label{eq:Lagrangian_general}
{\cal L} = \frac{F_0^2}{4} \text{Tr} (D_\nu \Sigma D^\nu \Sigma^\dagger) + \frac{F_0^2}{4} \text{Tr} (X \Sigma^\dagger + \Sigma X^\dagger )\,,
\ee
where $\Sigma$ describes the meson fields, $X=2 B_0( s + i p)$ represents scalar and pseudoscalar external fields and the covariant derivative is defined as 
\be
D_\mu \Sigma = \partial_\mu\Sigma - \frac{i}{2} [v_\mu,\Sigma] - \frac{i}{2} \{a_\mu,\Sigma \}\,,
\ee
where  $v_\mu$ and  $a_\mu$  are vectorial and axial external currents, respectively. The  Lagrangian has two free parameters $F_0$ and $B_0$. Because this is an effective theory, they have to be fixed by mesons phenomenology: they are related to the pion decay and to the quark-antiquark condensate, respectively. See for example~\cite{Gasser:1983yg, Leutwyler:1993iq, Ecker:1994gg,  Scherer:2002tk, Scherer:2005ri}. 

 If the meson field transforms as 
\be\label{eq:transSigma}
\Sigma \to R \Sigma L^\dagger\,,
\ee
the Lagrangian density possess the symmetry $SU(N_f)_L \times SU(N_f)_R $ 
and the chiral symmetry breaking corresponds to the spontaneous global symmetry breaking $SU(N_f)_L \times SU(N_f)_R \to SU(N_f)_{L+R}  $. In standard $\chi$PT, the mass eigenstates are charge eigenstates as well. This ensures that  mesons are particles with a well defined mass  and charge. If the system is in a medium the ground state can be tilted to different vacua that can also be charged with respect to some of the $SU(3)$ generators. This would imply that the eigenstates will not be eigenstates of the broken charges.
 The effects of a medium can be taken into account by considering appropriate external currents in Eq.~\eqref{eq:Lagrangian_general}. 

At vanishing temperature the ground state is determined by maximizing  the Lagrangian density with respect to the external currents. 
The pseudoscalar mesons are then described as oscillations around the vacuum. We use the same nonlinear representation of~\cite{Kogut:2001id} corresponding to 
\be\label{eq:sigma}
\Sigma= u \bar \Sigma u \qquad \text{with} \qquad u=e^{i  T \cdot \phi/2} \,,
\ee
where $T_a$ are the $SU(N_f)$ generators and  $ \bar \Sigma $ is a generic $SU(N_f)$ matrix to be determined by maximizing the static Lagrangian. The reasoning behind the  above expression  is that under $SU(N_f)_L \times SU(N_f)_R $ mesons can be identified as the fluctuations of the vacuum as in Eq.~\eqref{eq:transSigma} with $\theta_a^R = - \theta_a^L = \phi_a $. 

In the following we will assume that $a_\mu=0$, $p=0$, $X=2 G M$, where $M$ is the $N_f \times N_f$ diagonal quark mass matrix and $G$ is a constant, that with these conventions is equal to $B_0$. 
Moreover, we will assume that 
\be \label{ext}
v^\nu =  2 \mu \delta^{\nu 0} 
\ee
 meaning that the vectorial current consists of the quark chemical potential, with $\mu$  a $SU(N_f)$  matrix in flavor space. Its explicit expression in our case of interest, ie $SU(3)$, is:

\begin{align}
\mu=\text{diag}\left(\mu_u,\mu_d,\mu_s\right)= \text{diag}\left(\frac13 \mu_B+\frac12 \mu_I,\frac13 \mu_B-\frac12 \mu_I, \frac13
\mu_B-\mu_S\right)=\frac{\mu_B-\mu_S}3 I + \frac{\mu_I}2 \lambda_3 + \frac{\mu_S}{\sqrt{3}} \lambda_8\,,
\end{align}
It is important to remark that this model only holds for $|\m_B|\lesssim 940$ MeV, $|\m_I|\lesssim 770$ MeV and $|\m_S|\lesssim 550$ MeV \cite{Kogut:2001id}.

\subsection{Ground state and different phases} \label{ground}
\begin{figure}
\centering
\includegraphics[scale=0.5]{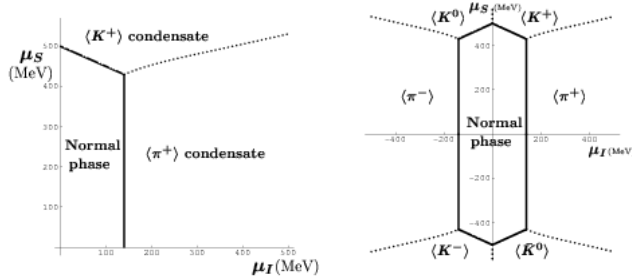}
\caption{Phase diagram of our model in presence of nonzero isospin and strangeness chemical potentials and at $T=0$. }
\label{phase}
\end{figure}

To find the ground state we have to substitute (\ref{eq:sigma}) in the Lagrangian (\ref{eq:Lagrangian_general}). It is not necessary to use a complete SU(3) parametrization for $\bar{\Sigma}$, but it is sufficient:
\begin{eqnarray}
\bar \Sigma=\left( \begin{array}{ccc} 1 & 0 & 0 \\
                                 0 & \cos \beta & -\sin \beta \\
                                 0 & \sin \beta & \cos \beta
                   \end{array} \right) \;
            \left( \begin{array}{ccc} \cos \alpha & \sin \alpha  & 0 \\
                                 -\sin \alpha & \cos \alpha & 0 \\
                                 0 & 0 & 1
                   \end{array} \right) \;
            \left( \begin{array}{ccc} 1 & 0 & 0 \\
                                 0 & \cos \beta & \sin \beta \\
                                 0 & -\sin \beta & \cos \beta
                   \end{array} \right),
\label{SadP}
\end{eqnarray}
because $\bar{\Sigma}$ has to be orthogonal to the chemical potential in the SU(3) generator space \cite{Mammarella:2015pxa}.\\
Substituting (\ref{SadP}) in (\ref{eq:Lagrangian_general}) and maximizing, we find three different vacua, thus there are three ground states,  each one describing a different phase \cite{Kogut:2001id}:
\begin{itemize}
\item Normal phase: \begin{align} \mu_I<m_\pi\, \qquad \qquad \mu_S<m_K-\frac12 \mu_I\,,\end{align}
 characterized by
\be
     \alpha_N=0,  \quad\beta_N \in (0,\pi), \quad \bar{\Sigma}_N=\text{diag}(1,1,1)\,.
\ee
\item Pion condensation phase: \begin{align} \mu_I>m_\pi\, \qquad \qquad \mu_S<
\frac{-m_\pi^2+\sqrt{(m_\pi^2-\mu_I^2)^2+4 m_K^2 \mu_I^2} }{2 \mu_I}\,,
\end{align}
 characterized by
$
    \cos \alpha_{\pi}=\left(\frac{m_\pi}{\mu_I}\right)^2,  \quad \beta_{\pi} =0\,$,  
\begin{align} \bar{\Sigma}_{\pi}=\left( \begin{array}{ccc}
                            \cos \alpha_\pi & \sin \alpha_\pi & 0 \\
                            -\sin \alpha_\pi & \cos \alpha_\pi & 0\\
                            0 & 0 & 1
                            \end{array} \right) = \frac{1+2 \cos\alpha_\pi}{3} I + i \lambda_2 \sin\alpha_\pi+ \frac{\cos\alpha_\pi-1}{\sqrt{3}}\lambda_8\,. \nonumber
                            \label{pion}
\end{align}
\item Kaon condensation phase: 
\begin{align}
\mu_S>m_K-\frac12 \mu_I\, \qquad
\mu_S>
\frac{-m_\pi^2+\sqrt{(m_\pi^2-\mu_I^2)^2+4 m_K^2 \mu_I^2 }}{2 \mu_I}\,,
\end{align}
characterized by
$
\cos \alpha_K=\left( \frac{m_K}{\frac12 \mu_I+\mu_S}\right)^2\,,  \beta_K=\pi/2\,$,
\begin{align} \label{kaon} \bar{\Sigma}_K
=\left( \begin{array}{ccc}
                            \cos \alpha & 0 & \sin \alpha  \\
                            0 & 1 & 0\\
                            -\sin \alpha & 0 & \cos \alpha
                            \end{array} \right)=
\frac{1+2 \cos\alpha_K}{3} I+ \frac{\cos\alpha_K-1}{2\sqrt{3}}\left(\sqrt{3}\lambda_3-\lambda_8\right) + i \lambda_5 \sin\alpha_K\,. \nonumber 
\end{align}
Note that the kaon condensation  can only happen for
\be \mu_S > \bar\mu_S=m_K - \frac{m_\pi}2\,.\ee 

\end{itemize}
These results are summarized in figure \ref{phase} \cite{Kogut:2001id}.

\section{Meson Masses and Mixing} \label{mm}
\subsection{Mixing}
As we have seen in Sec. \ref{ground} the ground state in the condensed phases is not diagonal and thus has SU(3) charges that cause symmetry breaking that will lead to mix mesons. Studying the breaking pattern can guide us to identify the allowed mixing in the condensed phases.

The starting Lagrangian has an $SU(3)_L \times SU(3)_R$ symmetry, broken to $SU(3)_V$ by the quark masses. The introduction of chemical potentials via the external current (\ref{ext}) further reduces this symmetry to $U(1)_{L+R}\times U(1)_{L+R}$, meaning that the $\l_3$ and $\l_8$ terms in (\ref{ext}) break isospin and hypercharge conservation. When the system enters one of the condensed phases the vacuum acquires a charge and thus there is no symmetry left.

In all of these cases we can use the quantum numbers of the $SU(2)$ subgroups of $SU(3)$ to label the states and to find out which ones can mix. We only need two of them, because they are not independent. The states that can mix and the related quantum numbers are shown in table \ref{table:mixing}

\begin{table}[h!]
\begin{center}
\begin{tabular}{|c|c|}\hline
Mixing states & $(T,U)$\\ \hline
$\pi_+,\pi_-$ & $(1,1/2)$ \\ \hline
$K_+,K_-$ & $(1/2,1/2)$ \\ \hline
$K_0,\bar K_0$ & $(1/2,1)$ \\
\hline
\end{tabular}
\end{center}
\caption{Mixing mesons  with the corresponding $T$-spin and $U$-spin quantum numbers. These quantum numbers label the $SU(3)$ subspace spanned by the corresponding mesonic states.  The $\pi_0$ and the $\eta$ do not appear because they are not $U$-spin eigenstates.}
\label{table:mixing}
\end{table}%

Unfortunately, this is not sufficient to determine if $\pi_0$ and $\eta$ can mix, because they do not have a well defined U and V spin, so we have to study deeply how the ground state affects them.
Let us first consider the normal phase. In the normal phase there is no operator that can induce the mixing of the mesonic  states, thus they remain unchanged but the $Q_3$ and $Q_8$ charges will induce Zeeman-like mass splittings.

In any of the condensed phases, there is an additional charge that is spontaneously induced, and the corresponding operator will lead to mixing.  

Let us first focus on isospin (or $T$-spin). We have to consider two cases.  Suppose that the vacuum has a charge that  commutes with $T^2$, as in the $\pi c$ phase,  say the charge corresponding to $T_2 = i( T_- - T_+)$, see Eq.~\eqref{pion}.  The $T_\pm$ operators can induce mixing among the charged pions and among the kaons. On the other hand, $T$-spin conservation does not allow the $\vert\pi_0\rangle =  \vert T=1, T_3=0\rangle$  to mix with the $\vert\eta\rangle = \vert T=0, T_3=0\rangle$.  

Now suppose instead that the vacuum has a charge that  does not commute with $T^2$ as in the $Kc$ phase. Any operator that does not commute with isospin will commute with $U$-spin or with $V$-spin. In the $K c$ phase  $Q_5 \vert 0 \rangle \neq 0$, then the vacuum is not invariant under this charge. However, since  $[T_5,U]=0$  it follows that $U$-spin is conserved. The lowering and raising operators inducing the mixing will be $U_\pm$. Regarding the $\pi_0$ and the $\eta$, in this case  we have that $\vert U=1, U_3=0\rangle$ and $\vert U=0, U_3=0\rangle$ do not mix. Since $\vert U=1, U_3=0\rangle = \frac{\vert\pi_0\rangle + \sqrt{3}\vert \eta\rangle}{2}$ and $\vert U=0, U_3=0\rangle = \frac{\sqrt{3}\vert  \pi_0\rangle -\vert \eta\rangle}{2}$, these will be the mass eigenstates.   


\subsection{Masses}
\begin{figure}[th!]
\includegraphics[width=7cm]{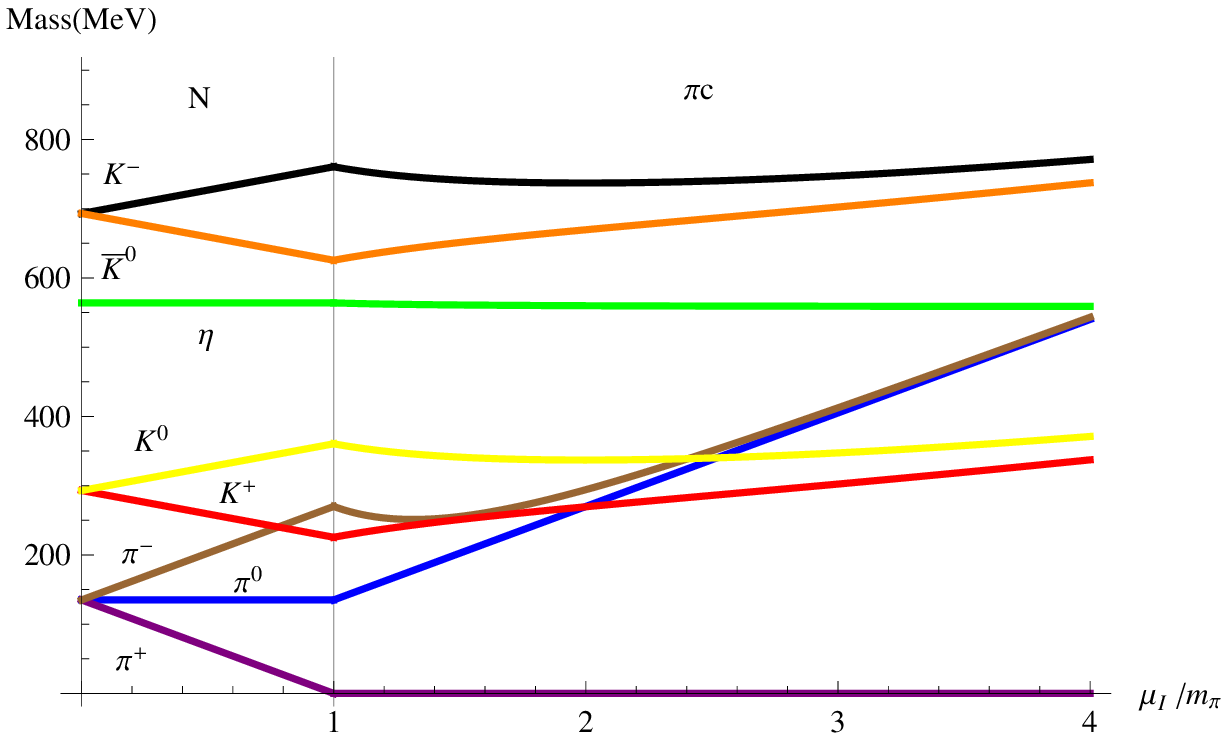}
\includegraphics[width=7cm]{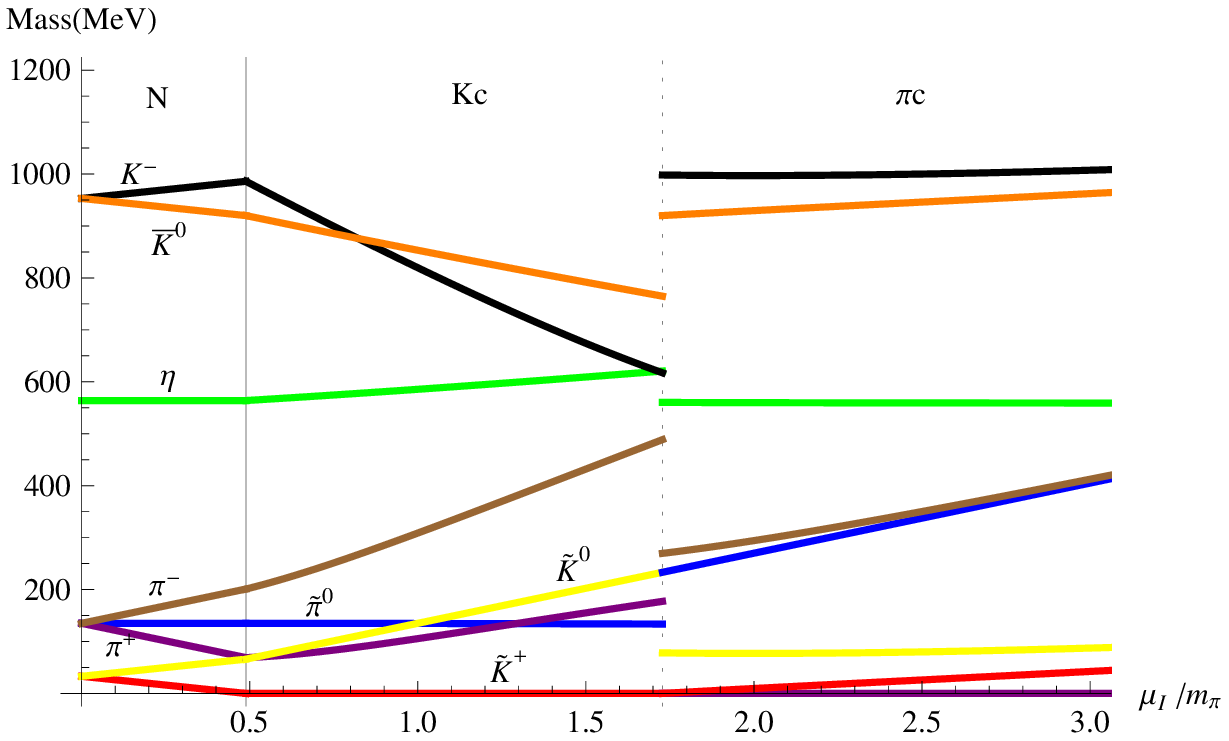}
\includegraphics[width=7cm]{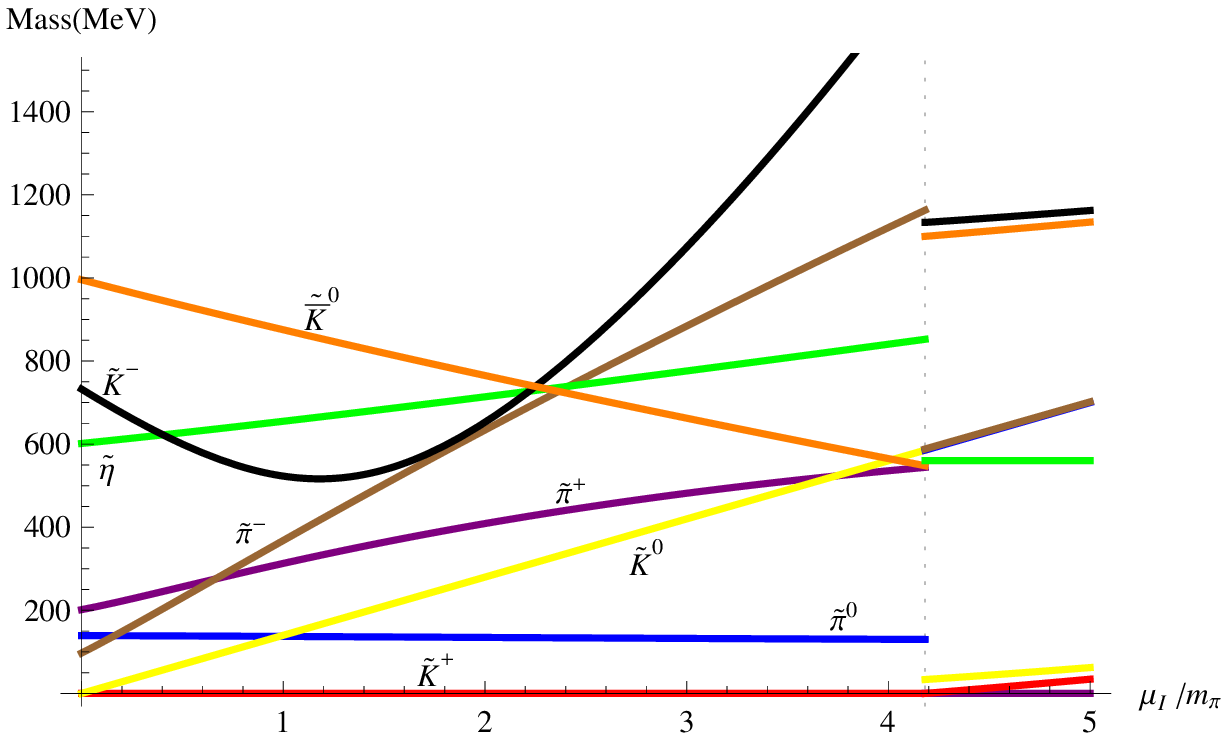}
\caption{(color online) Mass spectrum of the pseudoscalar mesonic octet.   Top left panel: results obtained  for $\mu_S=200$ MeV. The vertical solid line represents the second order phase transition between the normal phase and the pion condensation phase.  In this case the strange quark chemical potential is below the threshold value for kaon condensation, $425$ MeV, thus the kaon condensed phase does not take place for any value of $\mu_I$.  Top right panel: results obtained  for  $\mu_S=460$ MeV. The vertical solid line represents the second order phase transition from the normal phase to the kaon condensation phase. The dashed line corresponds to the first order phase transition between the kaon condensed phase to the pion condensed phase. Bottom panel: results obtained  for  $\mu_S=550$ MeV, corresponding to the largest value of $\mu_S$.}
\label{fig:masses}
\end{figure}

The mass eigenstates are found diagonalizing the Lagrangian in the different phases. They present the mixing predicted by the group theory analysis of the previous subsection. Masses and mixings have been calculated \cite{Mammarella:2015pxa}. The procedure is straightforward: we have to diagonalize the quadratic part of the Lagrangian by solving the related secular equation, obtaining the masses and subsequently the eigenstates. The resulting expressions are quite involved so we only show plots of the masses $\mu_I$ dependence for different values of $\mu_S$ in figure \ref{fig:masses}.

\section{Thermodynamic} \label{term}
\begin{figure}[t!]
\centering
\includegraphics[width=.4\textwidth]{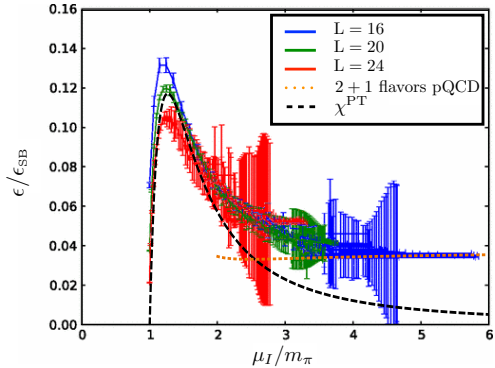}
\caption{(color online) Energy density over the Stefan-Boltzmann limit. The lattice points have been obtained at $T=20$ MeV; the  colors correspond to the different lattice volumes considered in~\cite{Detmold:2012wc}. The orange dotted line corresponds to the pQCD results of~\cite{Graf:2015pyl}. The dashed black line corresponds to our results.  }
\label{fig:energy_density}
\end{figure}

The pressure of the system can be found substituting $\bar{\Sigma} \rightarrow \Sigma$ in \ref{eq:Lagrangian_general}. In the different phases we get:
\begin{align}
p^N_\text{LO}=0 \qquad p^{\pi c}_\text{LO} = \frac{F_0^2\mu_I^2}{2} \left(1- \frac{m_\pi^2}{\mu_I^2}\right)^2 \, \qquad p^{Kc}_\text{LO} = \frac{F_0^2\mu_K^2}{2} \left(1- \frac{m_K^2}{\mu_K^2}\right)^2 \label{eq:pressure-pions} \nn
\end{align} with $\mu_K=\mu_I/2+\mu_S$. It is important to observe that the expressions for the pressure in the condensed phases are exactly the same if we change $\mu_I \rightarrow \mu_K$ and $m_\pi \rightarrow m_K$.
Number density and equation of state can be derived using:
\be
n_{i, \text{LO}}=\frac{\partial p_\text{LO}}{\partial \mu_i} \qquad \text{and} \qquad \epsilon=  \mu_I n_I + \mu_S n_S - p\,,
\label{def:nI}
\ee where the index $i$ represents I=isospin or K=kaon.
In the $\pi C$ phase we obtain:
\be
n_{I, \text{LO}}^{\pi C}=f_\pi^2 \mu_I \(1-\frac{m_\pi^4}{\mu_I^4}\)\, \qquad \text{and} \qquad \epsilon^{\pi c}_\text{LO}=\frac{f_\pi^2  \mu_I^2}{2}\left(1+2\frac{ m_\pi^2}{\mu_I^2}  - 3 \frac{m_\pi^4}{\mu_I^4}\right)\,
\label{eq:nI}
\ee while the analogous results in the kaon condensation phase can be obtained substituting $m_\pi \to m_K$ and $\mu_I \to \mu_K$, see \cite{Detmold:2008yn,Carignano:2016rvs}.
To compare  the $\chi$PT energy density  with the results obtained by LQCD simulations in~\cite{Detmold:2012wc} and by pQCD in~\cite{Graf:2015pyl},  we divide it by the Stefan-Boltzmann limit, which has been defined in~\cite{Detmold:2012wc}   as $ \epsilon_{SB}=9 \m_I^4/(4 \pi^2) $ and describes a cold, degenerate gas of weakly interacting quarks which is the limit of our system if there was no interactions. In Fig.~\ref{fig:energy_density} we report our ratio $\epsilon^{\pi c}_\text{LO}/\epsilon_\text{SB}$ and the results of~\cite{Detmold:2012wc} and~\cite{Graf:2015pyl}. We immediately notice that the $\chi$PT curve perfectly captures the peak structure at low $\mu_I$, while it begins to depart from the LQCD  results after $\m_I \sim 2 m_{\pi}$, indicating the breakdown of the LO approximation. An interesting result is that within our framework we obtain an analytic expression for the position of the peak, which for the $\pi c$ phase is given by $\mu_I^\text{peak} = \({\sqrt{13} -2}\, \)^{1/2} m_\pi \simeq 1.276\, m_\pi $
and is independent of $f_\pi$. This result is very close to the LQCD result obtained in~\cite{Detmold:2012wc}, where the values $ \mu_I^\text{peak} = \{1.20,1.25,1.275\} m_\pi  $  have been obtained considering  different spatial volumes $L^3$ with side   $L=\{16,20,24\}$, respectively. The  continuum-linearly-extrapolated value for the peak position 
 is $ \mu_I^\text{peak} = 1.30~ m_\pi$. 

\begin{figure}[t!]
\centering
\includegraphics[width=.4\textwidth]{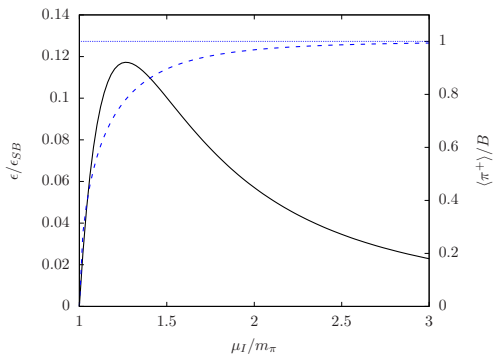}
\caption{(Color online) $\chi$PT results for the  ratio $\epsilon/\e_{SB}$ (solid line) and the pion condensate normalized to its maximum value $B$ (dashed blue line). }
\label{fig:pioncond}
\end{figure}

It is not clear if the peak point is related to some undergoing physics. In our model the pion
condensate has reached a significant value at that point, almost
approaching its asymptotic limit $B$, see
Fig.~\ref{fig:pioncond}, so we tend to exclude that the peak appearance is related to the occurrence of the condensation.\\
More specifically, the pion condensate can be evaluated by introducing external source terms 
in the chiral Lagrangian and by differentiating with respect to them. After the phase transition it grows as 
$\langle \pi^+ \rangle = B \sqrt{1-{m_\pi^4}/{\mu_I^4 }}$~\cite{Kogut:2001id, Kogut:2000ek}, 
so that at $\mu_I^{\text{peak}}$ it reaches already $\sim 79\%$ of its asymptotic value.
 We are therefore more inclined to interpret this peak structure as a consequence of the filling of the condensate, rather
than of the onset of its formation, which at zero-temperature
should occur  at $\mu_I = m_\pi$.  We also obtain an analytic expression for the ratio at the peak
\be
\left.\frac{\epsilon}{\epsilon_{SB}}\right\vert_\text{peak} =\frac{4 (\sqrt{13}-5) \pi^2 }{9(\sqrt{13}-2)}\frac{f_\pi^2}{m_\pi^2}\,,
\label{eq:enpeak}
\ee
which would give information on the $f_\pi/m_\pi$ scaling if precise LQCD data  were available.
It is  worth mentioning that the above results hold in the $Kc$ phase if one considers 
$ \epsilon_{SB}=9 \mu_K^4/(4 \pi^2) $. Then, we obtain $\mu_K^\text{peak} = (\sqrt{13} -2)^{1/2} m_K$  and an expression analogous to Eq.\eqref{eq:enpeak}, with $m_\pi \to m_K$.

From Fig.~\ref{fig:energy_density} it seems that for $\mu_I > 2 m_\pi$ the LO $\chi$PT breaks down, resulting in an underestimate of  the energy density. The basic reason is that for large $\mu_I$ we have  that $\epsilon^{\pi c}_\text{LO} \propto f_\pi^2 \mu_I^2$ from Eq.~\eqref{def:nI}, while pQCD correctly predicts  $\epsilon^{\pi c} \propto  \mu_I^4$.     At the  next-to-leading order (NLO), it is possible to show that the $\chi$PT energy density  includes a term proportional to  $(2 l_1+2l_2 + l_3) \mu_I^4$, where $l_1$, $l_2$ and $l_3$ are NLO low energy constants~\cite{Scherer:2002tk}.  Comparing our results with the pQCD energy density we obtain \be\label{eq:LEC} 2 l_1+2l_2 + l_3 = \frac{3}{8\pi^2}\left(\frac{\epsilon_\text{pQCD}}{\epsilon_\text{SB}} - \frac{2 f_\pi^2 \pi^2 }{9\mu_I^2}  + {\cal O}(\mu_I^{-4})\right)  \,,\ee leading to $2 l_1+2l_2 + l_3 \simeq0.6 \times 10^{-3}$ for $\mu_I= 3 m_\pi$, consistent with the empirical values of the low energy costants ~\cite{Scherer:2002tk}.

\section{Conclusion} \label{conc}
We have shown how meson physics in presence of chemical potentials can be described using Chiral Perturbation Theory. There are three different phases: a normal phase, a pion condensation phase and a kaon condensation phase. The condensed phases ground states have a charge and thus can generate mixing among mesons.\\
In this context we have illustrated how the mixing is influenced by model symmetries and that group theory constrains the mixing possibilities. These results, obtained by group theory alone, are expected to hold in any theory describing meson states. We have then described mesons masses in the three phases, calculating them from the Lagrangian (\ref{eq:Lagrangian_general}). These masses are in perfect agreement with the group theory reasoning.\\
Finally we have derived the pressure and equation of state of the system at $T=0$, comparing them with the equivalent results obtained in lattice simulations and obtaining a very good agreement on a non trivial feature like a peak structure in the meson energy density over the Stefan Boltzmann one.\\
These results can be applied for example in the physics of compact stars, the study of cosmic rays and the study of in medium nuclear decays.\\

\end{document}